**Carbon nanotube, graphene, nanowire, and molecule-based electron and spin transport phenomena using the non-equilibrium Green's function method at the level of first principles theory**

Woo Youn Kim and Kwang S. Kim[*]

*Center for Superfunctional Materials, Department of Chemistry, Pohang University of Science and Technology, San 31, Hyojadong, Namgu, Pohang 790-784, Korea*

[*]kim@postech.ac.kr

**Abstract:**
Based on density functional theory (DFT), we have developed algorithms and a program code to investigate the electron transport characteristics for a variety of nanometer scaled devices in the presence of an external bias voltage. We employed basis sets comprised of linear combinations of numerical type atomic orbitals and k-point sampling for the realistic modeling of the bulk electrode. The scheme coupled with the matrix version of the non-equilibrium Green's function method enables determination of the transmission coefficients at a given energy and voltage in a self-consistent manner, as well as the corresponding current-voltage (*I-V*) characteristics. This scheme has advantages because it is applicable to large systems, easily transportable to different types of quantum chemistry packages, and extendable to describe time-dependent phenomena or inelastic scatterings. It has been applied to diverse types of practical electronic devices such as carbon nanotubes, graphene nano-ribbons, metallic nanowires, and molecular electronic devices. The quantum conductance phenomena for systems involving quantum point contacts and *I-V* curves are described for the dithiol-benzene molecule in contact with two Au electrodes using the k-point sampling method.

**I. Introduction**

One of the most rapidly rising fields in physics or chemistry today is nanoelectronics. A recent advance in nanotechnology has shown possibility of the realization of nanometer-scaled devices.[1-14] The nanoelectronics, which is a common designation of nanometer-scaled electronics including molecular electronics, spintronics, and carbon-based electronics, have been considered as a promising device complementing the conventional silicon based electronics.[15, 16] In order to overcome the fundamental limitations of the silicon-based electronics, the study of the nanoelectronics has to (i) develop all the conventional electronic components such as wires, diodes, transistors, and memory devices based on single molecules and (ii) devise a new conceptual electronic device based purely on quantum effects like quantum computing devices. To this end, various types of materials and devices have been studied.[10, 17-22] For example, conductance of single molecules through nano-scaled junctions has been measured.[2, 5, 7-10, 17] Carbon nanotubes (CNTs) have been intensively investigated as a candidate of the next generation of field-effect transistor.[6, 19, 23] Recently, graphenes are rising as the most promising material with a variety of applications.[20, 22, 24, 25] However, technical problems in fabrication process of such nanodevices hinder the progress in the field.[16, 26] To advance in this field further, more concrete theoretical understanding is essential.

In this regard, we have developed a theoretical tool to explain and predict electron transport phenomena through diverse types of nanometer-scaled devices. Our scheme is based on the DFT coupled with non-equilibrium Green's function (NEGF) method.[27, 28] Despite that most semi-empirical and first principles calculations in principle has not been developed for the description of electronic structures of systems under non-equilibrium conditions, its validity under such conditions has been shown by various studies.[29-42] Some of them were based on tight-binding or semi-empirical methods[34, 37, 38] and few were based on first-principle method with full description of electron density converged self-consistently under non-equilibrium conditions which treat the whole system on the same footing.[29, 40, 41] In this scheme, the role of the NEGF method is to describe the electronic structures of open systems composed of two bulk electrode parts and a device part. To accurately describe semi-infinite nature of bulk surface, the realistic modeling of the electrode parts is essential.[36] In order to achieve such realistic modeling, efficient memory management is important.

For this purpose, the use of the Spanish Initiative for Electronic Simulations with Thousands of Atoms (SIESTA) program package[43] would be one of practical approaches. Since SIESTA uses numerical atomic orbital basis sets and pseudo-potentials, it is suitable to handle large systems. This approach has shown its accuracy and efficiency by successfully applying to a variety of different chemical or biological systems in equilibrium states.[44-48] By implementing the NEGF method, we have extended the SIESTA code to investigate the transport phenomena for nanometer-size systems in non-equilibrium states. There have been two independent programs.[29, 40] For instance, the TranSIESTA code was developed first for non-spin-polarized transport phenomena, while recently the option of spin-polarized transport and k-point sampling are available but the algorithm employed has not been reported. The SMEAGOL code is able to do such calculations, but we note that it is not practical for calculating large systems. Our new code enables to do all of them. In particular, our version is highly flexible to be upgraded because the NEGF part is written independently from the main DFT code. As a result, it is easy to implement the code in other computational methods such as Hartree-Fock, DFT, configuration interaction, tight-binding, or other high levels of *ab initio* theory. Here, to facilitate our discussion, we show how to implement the NEGF method to a general DFT code (not just for the SIESTA), and particularly demonstrate some important examples for nanoelectronic devices using our code.

The organization of the paper is as follows: First, we discuss the systematic set up of the scattering problem for explicit electron/spin transport calculations for a given electronic device. Second, we describe the algorithms used in our scheme with detailed formalism. Third, we explain how to calculate density matrix for open systems using the matrix version of the Green's function (MGF) method together with the self-energy matrix in the equilibrium condition and its extension form to describe the non-equilibrium states. Finally, we show our results on systems such as CNTs with dopants, graphene nano-ribbons, gold (Au) and nickel (Ni) nanowires (NWs), and the dithiol bezene molecule linked by Au bulk electrodes at both ends.

## II. General Formalism and Implementation

The NEGF formalism, also named as Keldysh formalism, is the representative method for describing non-equilibrium states, and its practical usefulness has been discussed in literature.[27, 28] Coupled with the Landauer-Büttiker formalism, it provides a general approach to describe quantum transport phenomena including interactions such as electron-electron or electron-phonon interaction. Since the details of derivation and meaning of the formalism are already well known, we focus on implementing them in general DFT codes. To avoid complexity, all the formula in this paper are based on a one particle picture similar to that used in literature.[27]

### A. Set up of the Device Structure

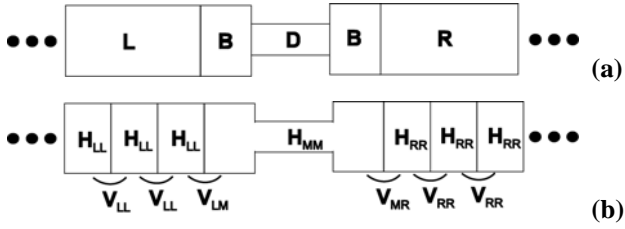

**Figure 1. (a)** A schematic structure of usual device composed of left/right bulk electrodes (L/R) and a device part (D) contacted between two buffer regions (**B**'s) which screen the interaction between D and L/R. **(b)** Structure of the device represented in terms of Hamiltonian matrix elements.

Figure 1 shows a schematic structure of usual devices composed of the left/right bulk electrode parts (L/R) and the device part (D) which involves the scattering region contacted between them. Since each bulk electrode is a semi-infinite system, it is impossible to handle the system directly. The potential arisen from the scattering region does not influence the bulky region of electrodes, because both electrodes are metallic and they can effectively screen external potentials within a few layers from the surface. Considering the screening effect, we can set up an effective system of a finite dimension which can accurately describe the electronic structure of the device part. To this end, we introduce the self-energy term for each electrode and the buffer regions (**B**'s) connecting between the device part and the bulk electrode parts, as shown in Figure 1. The self-energy involves all the information of semi-infinite nature of the bulk electrodes. The buffer regions block the interaction between the device part and the electrode parts by screening the potential induced from one another.

If we describe the above discussion in terms of Hamiltonian, the introduction of the self-energy and buffer region reduces the dimension of the Hamiltonian matrix from the infinite to a finite one:

$$H^{k_\|} = \begin{bmatrix} \ddots & \vdots & & \vdots & & & \\ & H_{LL}^{k_\|} & V_{LL}^{k_\|} & 0 & 0 & 0 & \\ & V_{LL}^{k_\|\dagger} & H_{LL}^{k_\|} & H_{LM}^{k_\|} & 0 & 0 & \\ \cdots & 0 & H_{ML}^{k_\|} & H_{MM}^{k_\|} & H_{MR}^{k_\|} & 0 & \cdots \\ & 0 & 0 & H_{RM}^{k_\|} & H_{RR}^{k_\|} & V_{RR}^{k_\|} & \\ & 0 & 0 & 0 & V_{RR}^{k_\|\dagger} & H_{RR}^{k_\|} & \\ & & \vdots & & \vdots & & \ddots \end{bmatrix} \xrightarrow{\text{Screening approximation}} H_{eff}^{k_\|} = \begin{bmatrix} H_{LL}^{k_\|} + \Sigma_L^{k_\|} & H_{LM}^{k_\|} & 0 \\ H_{ML}^{k_\|} & H_{MM}^{k_\|} & H_{MR}^{k_\|} \\ 0 & H_{RM}^{k_\|} & H_{RR}^{k_\|} + \Sigma_R^{k_\|} \end{bmatrix}_{N \times N} \quad (1)$$

Lower labels of each term in the Hamiltonian matrix elements represent regions to which the given basis orbital belongs, and $k_{//}$ is a reciprocal lattice vector point along a surface-parallel direction (orthogonal to the transmission direction) in the irreducible Brillouin zone (IBZ).

In order to derive the explicit expression of the self-energy terms, we use the MGF approach.[27] The MGF for the device of both electrodes is given by

$$\begin{bmatrix} E\bar{S}_{LL}^{k_\|} - \bar{H}_{LL}^{k_\|} & E\bar{S}_{LD}^{k_\|} - \bar{H}_{LD}^{k_\|} & 0 \\ E\bar{S}_{DL}^{k_\|} - \bar{H}_{DL}^{k_\|} & ES_{DD}^{k_\|} - H_{DD}^{k_\|} & E\bar{S}_{DR}^{k_\|} - \bar{H}_{DR}^{k_\|} \\ 0 & E\bar{S}_{RD}^{k_\|} - \bar{H}_{RD}^{k_\|} & E\bar{S}_{RR}^{k_\|} - \bar{H}_{RR}^{k_\|} \end{bmatrix} \begin{bmatrix} \bar{G}_{LL}^{k_\|}(E) & \bar{G}_{LD}^{k_\|}(E) & \bar{G}_{LR}^{k_\|}(E) \\ \bar{G}_{DL}^{k_\|}(E) & G_{DD}^{k_\|}(E) & \bar{G}_{DR}^{k_\|}(E) \\ \bar{G}_{RL}^{k_\|}(E) & \bar{G}_{RD}^{k_\|}(E) & \bar{G}_{RR}^{k_\|}(E) \end{bmatrix} = I^{k_\|} \quad (2)$$

The lower label "$_D$" represents the region of the device part including the buffer region and unit cells of both electrodes (L-B-D-B-R). $H_{DD}^{k_\|}$ is expressed as

$$H_{DD}^{k_\|} = \begin{bmatrix} H_{LL}^{k_\|} & H_{LM}^{k_\|} & 0 \\ H_{ML}^{k_\|} & H_{MM}^{k_\|} & H_{MR}^{k_\|} \\ 0 & H_{RM}^{k_\|} & H_{RR}^{k_\|} \end{bmatrix}.$$

The upper bar of each block matrices ($\bar{S}, \bar{H}, \bar{G}$) denotes that its dimension is semi-infinite, as shown in the following examples:

$$\bar{H}_{RR}^{k_\|} = \begin{bmatrix} H_{RR}^{k_\|} & V_{RR}^{k_\|} & 0 & \cdots \\ V_{RR}^{k_\|\dagger} & H_{RR}^{k_\|} & V_{RR}^{k_\|} & \\ 0 & V_{RR}^{k_\|\dagger} & H_{RR}^{k_\|} & \\ \vdots & & & \ddots \end{bmatrix}$$

Simple matrix algebra gives the following solution for the Green's function of the device part [$G_{DD}^{k_\|}(E)$] in Eq. 2 with the expression of the self-energy terms:[27, 28]

$$G_{DD}^{k_\|}(E) = \left[ ES_{DD}^{k_\|} - H_{DD}^{k_\|} - \Sigma_L^{k_\|}(E) - \Sigma_R^{k_\|}(E) \right]^{-1} \quad (3)$$

$$\Sigma_L^{k_\|}(E) = \left[ E\bar{S}_{DL}^{k_\|} - \bar{H}_{DL}^{k_\|} \right] G_{LL}^{k_\| \, surf}(E) \left[ E\bar{S}_{LD}^{k_\|} - \bar{H}_{LD}^{k_\|} \right] \quad (4\text{-}1)$$

$$\Sigma_R^{k_\|}(E) = \left[ E\bar{S}_{DR}^{k_\|} - \bar{H}_{DR}^{k_\|} \right] G_{RR}^{k_\| \, surf}(E) \left[ E\bar{S}_{RD}^{k_\|} - \bar{H}_{RD}^{k_\|} \right] \quad (4\text{-}2)$$

where we define

$$G_{00}^{k_\| \, surf}(E) = \left[ ES_{00}^{k_\|} - H_{00}^{k_\|} \right]^{-1}.$$

Note that although each self-energy matrix is obtained by multiplying three matrices of infinite or semi-infinite dimensions, the dimension of the result is finite and it is the same with that of $H_{DD}^{k_\parallel}$. The effective Hamiltonian used in Eq. 1 is defined as follows.

$$H_{eff}^{k_\parallel} \equiv H_{DD}^{k_\parallel} - \Sigma_L^{k_\parallel}(E) - \Sigma_R^{k_\parallel}(E) \qquad (5)$$

From now on, we omit the lower label "$_{DD}$" for convenience's sake.

**B. Self-Energy and Surface Green's Function**

To obtain the self-energy matrix for the electrodes, we need to calculate the bulk systems corresponding to the electrode parts separately from the L-B-D-B-R calculation. Our strategy is the following. The first step is to calculate the Green's function of the semi-infinite electrodes. We divide the unit cell of the periodic system into several layers so that atoms in one unit cell solely interact with atoms in the nearest neighbor unit cell as presented in Figure 2.

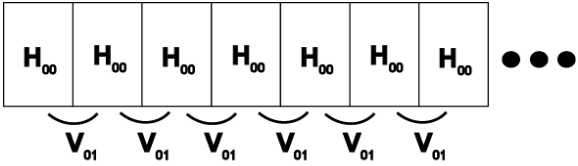

**Figure 2.** A schematic structure of the bulk electrodes in terms of Hamiltonian matrix elements.

Eq. 6 gives the Green's function equation for the system.

$$\begin{bmatrix} ES_{00}^{k_\parallel}-H_{00}^{k_\parallel} & ES_{01}^{k_\parallel}-V_{01}^{k_\parallel} & 0 & \cdots \\ ES_{10}^{k_\parallel}-V_{10}^{k_\parallel} & ES_{00}^{k_\parallel}-H_{00}^{k_\parallel} & ES_{01}^{k_\parallel}-V_{01}^{k_\parallel} & \\ 0 & ES_{10}^{k_\parallel}-V_{10}^{k_\parallel} & ES_{00}^{k_\parallel}-H_{00}^{k_\parallel} & \\ \vdots & & & \ddots \end{bmatrix} \begin{bmatrix} G_{00}^{k_\parallel} & G_{01}^{k_\parallel} & G_{02}^{k_\parallel} & \cdots \\ G_{10}^{k_\parallel} & G_{00}^{k_\parallel} & G_{12}^{k_\parallel} & \\ G_{20}^{k_\parallel} & G_{21}^{k_\parallel} & G_{00}^{k_\parallel} & \\ \vdots & & & \ddots \end{bmatrix} = \begin{bmatrix} I_{00}^{k_\parallel} & 0 & 0 & \cdots \\ 0 & I_{00}^{k_\parallel} & 0 & \\ 0 & 0 & I_{00}^{k_\parallel} & \\ \vdots & & & \ddots \end{bmatrix} \qquad (6)$$

Here, $H^{k_\parallel}$ and $S^{k_\parallel}$ are calculated by one-dimensional Fourier transform of $H^k$ and $S^k$ obtained from the 3D bulk calculation along the surface normal direction (parallel to the transmission directon) $k_\perp$, as follows:

$$H_{0n}^{k_\parallel} = \frac{1}{N_{k_\perp}} \sum_{k_\perp} e^{-ik_\perp \cdot R_n} H^{k_\parallel, k_\perp},$$

$$S_{0n}^{k_\parallel} = \frac{1}{N_{k_\perp}} \sum_{k_\perp} e^{-ik_\perp \cdot R_n} S^{k_\parallel, k_\perp}.$$

where $R_n$ is the lattice vector of surface-normal direction of n-th layer, and $N_{k_\perp}$ is the number of $k_\perp$ points in the IBZ. The dimension of the matrix is again infinite, but each block in the matrix repeats according to the periodicity of the unit cell. Therefore, we only need to know the Green's function for one unit cell to extract the information of the surface properties. The function we need is called 'surface Green's function ($G^{surf}$)'. It can be calculated recursively as follows.

$$G_{00}^{k_\parallel}(E) = \left[ ES_{00}^{k_\parallel} - H_{00}^{k_\parallel} - \left\{ ES_{01}^{k_\parallel} - V_{01}^{k_\parallel} \right\} G_{00}^{k_\parallel}(E) \left\{ ES_{10}^{k_\parallel} - V_{10}^{k_\parallel} \right\} \right]^{-1} \quad (7)$$

In our code, we use the transfer matrices method which shows much faster convergence rate than the recursion method; the transfer matrices method converges at the rate of order of $2^N$, as compared to the convergence rate of the order N for the recursion method.[49] To obtain well converged results, the latter needs to perform the inverting processes several hundred times, whereas the former requires only a few steps. Moreover, we must calculate the surface Green's function at each energy value $E$ on the given contour points, which we will discuss in the next section. Thus, the transfer matrix method enormously reduces the computer resources in this step. The second step is to generate the self-energy matrix from the surface Green's function using Eqs. 4-1,2. In both equations, the interaction terms multiplied on both sides of the surface Green's function are obtained from the calculation of the L-B-D-B-R system.

**C. Density Matrix in the Equilibrium**

Because we completed the effective Hamiltonian matrix by combining the self-energy matrix to the original Hamiltonian matrix as given in Eq. 5, we are ready to calculate the density matrix (DM) under the equilibrium condition ($DM_{eq}$). The $DM_{eq}$ is obtained by integrating the Green's function multiplied by the Fermi-Dirac distribution function along the real energy axis:[50]

$$DM_{eq}^{k_\parallel} = -\frac{1}{\pi} \text{Im} \int_{-\infty}^{\infty} G^{k_\parallel}(E) f(E-\mu) dE \quad (8)$$

For a wide range of the energy spectrum, it requires tremendous computational resources to perform accurate integration along the real energy domain. The equivalent result can be obtained by performing the integration along a certain contour on the imaginary plane. The idea is the following. The integrand of Eq. 8 can be transformed to another expression from the following two identities:

$$\oint G^{k_\parallel}(z) f(z-\mu) dz = 2\pi i \sum_{k=1}^{n} \underset{z=z_k}{\text{Res}} \left[ G^{k_\parallel}(z) f(z-\mu) \right] \quad (9)$$

where $z_k = i(2k+1)\pi k_B T$.

$$\int_{-\infty}^{\infty} G^{k_\parallel}(E) f(E-\mu) dE = -\int_C G^{k_\parallel}(z) f(z-\mu) dz - 2\pi i k_B T \sum_{k=1}^{n} G^{k_\parallel}(z_k) \quad (10)$$

In Eq. 9, $Z_k$ is a pole of the Fermi-Dirac distribution function, so called, Matsubara frequency. The first term on the left side in Eq. 10 is the integration along a certain contour C on the imaginary plane.

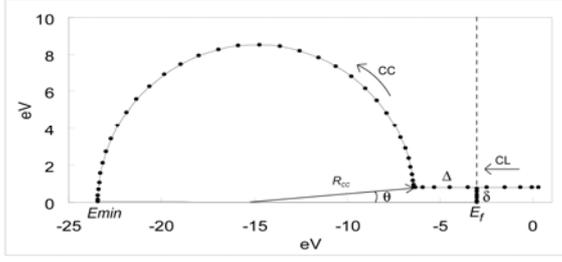

**Figure 3.** An example of the contour points on the imaginary plane, which is obtained by the Gaussian quadrature method. CL and CC represent the direction of the contour integral. $\Delta$, $\delta$, $R_{cc}$, and $\theta$ are parameters to determine the shape of the contour. $E_{min}$ is the minimum energy points on the contour. $E_f$ is the Fermi energy.

Figure 3 shows an example of such contour points. To get the integration points and the corresponding weight factors along the contour, we use the Gaussian quadrature method based on the Gauss-Legendre integration.[51] In this way, it requires only ~40 contour points to obtain a reasonable DM.

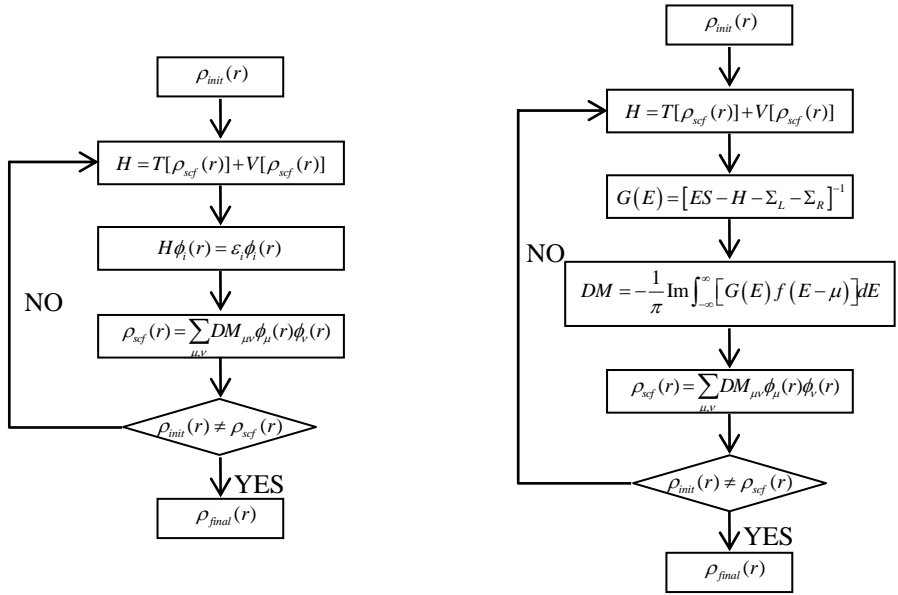

**Figure 4.** Algorithms of the DFT (left) and the DFT coupled with MGF (right).

In summary, because the system considered here is an open system contacted by two semi-infinite bulk systems, it is intractable to handle directly. By introducing the self-energy terms in the screening approximation, it becomes possible to calculate density matrix of the system. Coupling this scheme with DFT gives a converged electron density in a self-consistent manner. Figure 4 shows algorithms of the DFT method and the DFT-MGF coupled method. The difference between these two methods is that, in the formal method, the DM is obtained by diagonalizing the Hamiltonian matrix, while in the latter case, DM is obtained by integrating the Green's function along a certain contour on the imaginary plane.

**D. Effects of the Finite Bias Voltage**

Up to now, we have discussed the scheme to calculate DM for the open system under equilibrium condition. To generate finite current flow through the device, we should consider the system under external bias voltage which becomes a non-equilibrium state because of different chemical potentials of electrodes. In the non-equilibrium state, there are two important effects to consider. First, spatial distribution of the electron density in the scattering region becomes non-static. However, all the functional in DFT [e.g. local density approximation (LDA) or generalized gradient approximation (GGA)] has been established for the static electron density. Moreover, DFT is not appropriate to describe excited states where the excited states can be used to represent conduits for the electron transport.[52] Nonetheless, DFT is widely used because it is easy to be implemented with the NEGF method. Also, the incorporation of the GW method[53] or time-dependent DFT[54] method can further better represent the correlation effects induced in the scattering process for the next stage. Here, we assume that electron transport is in the steady state so that the distribution of the electron density becomes static such that the static functional (LDA or GGA) works in this particular system.

The second effect is the split of the chemical potentials of both electrodes due to the bias voltage. This gives rise to the change of the Hartree potential of the device part. We use the Poisson equation for correction of the electron static potential induced from the extra electron density [$\rho_{extra}(r) = \rho_{neq}(r) - \rho_{eq}(r)$, where $\rho_{neq}(r)$ is the electron density in the non-equilibrium and $\rho_{eq}(r)$ is the electron density in the equilibrium] as

$$\nabla^2 V^{eff}(r) = -4\pi \rho_{extra}(r) \qquad (11)$$

which has the following solution.[43]

$$V^{eff}(r) = \varphi(r) + \vec{c}_1 \cdot \vec{r} + c_2$$
$$= \varphi(r) - V\left(\frac{z}{L} - \frac{1}{2}\right)$$

Figure 5 presents the natural boundary condition of the Poisson equation along the junction of the system established by the bias voltage. The boundary condition gives two undetermined constants ($c_1$, $c_2$) in the solution of the second order differential equation, Eq. 11. Here, we assumed that in both electrode regions the potential drop does not arise because of the strong screening effect in the bulk metal. Therefore, the potential drop solely arises in the device part including the buffer regions (i.e. B-D-B region). Change of the chemical potential for each lead (L or R regions) can be achieved by constantly shifting the Hamiltonian matrix elements which is obtained from the independent calculation of the bulk electrodes.

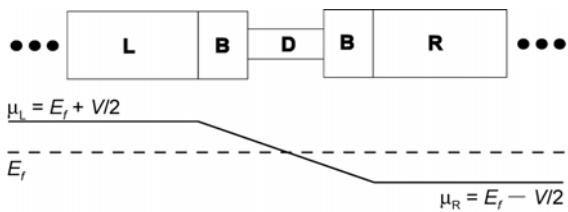

**Figure 5.** Potential drop through the junction due to the external bias voltage $V$. $\mu_{L/R}$ is the chemical potential of the left/and right electrodes, respectively and $E_f$ is the Fermi energy of the system.

Considering the above two effects, the DM under the non-equilibrium condition can be calculated using the lesser Green's function [$G^<(E)$] as follows.[27, 28]

$$DM^{k_\parallel} = \frac{1}{2\pi i}\int G^{<k_\parallel}(E)dE$$
$$= DM_{eq}^{Lk_\parallel} + \Delta DM_{neq}^{R\ k_\parallel} \quad (12\text{-}1)$$
$$= DM_{eq}^{Rk_\parallel} + \Delta DM_{neq}^{L\ k_\parallel} \quad (12\text{-}2)$$

where we define

$$G^{<k_\parallel}(E) = 2\pi i \left[ A_L^{k_\parallel}(E) f(E-\mu_L) + A_R^{k_\parallel}(E) f(E-\mu_R) \right]$$
$$DM_{eq}^{L/Rk_\parallel} = \int_{-\infty}^{\infty} A^{k_\parallel}(E) f(E-\mu_{L/R}) dE$$
$$\Delta DM_{neq}^{L/Rk_\parallel} = \int A_{L/R}^{k_\parallel}(E) \left[ f(E-\mu_{L/R}) - f(E-\mu_{L/R}) \right] dE.$$

Here, $A(E)$ and $A_{L/R}(E)$ are the total spectral function and the spectral function of the left/right junctions, respectively, and they are defined as follows.

$$A^{k_\parallel}(E) = A_L^{k_\parallel}(E) + A_R^{k_\parallel}(E)$$
$$A_{L/R}^{k_\parallel}(E) = \frac{1}{2\pi} G^{k_\parallel} \Gamma_{L/R}^{k_\parallel} G^{k_\parallel \dagger} \quad (13)$$
$$\Gamma_{L/R}^{k_\parallel}(E) = i\left[ \Sigma_{L/R}^{k_\parallel}(E) - \Sigma_{L/R}^{k_\parallel \dagger}(E) \right] \quad (14)$$

where $\Gamma_{L/R}^{k_\parallel}(E)$ are imaginary values of the self-energy of the left/right junctions, so called the gamma function. Thus, the DM is the sum of the equilibrium part ($DM_{eq}^{L/Rk_\parallel}$) and the extra part induced by the non-equilibrium ($\Delta DM_{neq}^{L/Rk_\parallel}$). The former is calculated by the contour integration technique as we discussed before, and the latter is obtained by the real space integration technique. In Eq. (12-1,2), the DM has two equivalent expressions which are the sum of the equilibrium part with left/right chemical potential and the extra non-equilibrium part. However, in the numerical integration, the two expressions give different results. We use the average value of two results with a certain weighting factor mixing two contributions, so called double contour technique. Figure 6 shows an example of the contour used in the double contour technique.

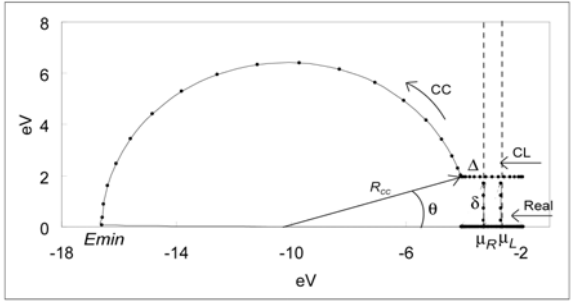

**Figure 6.** An example of the contour points used in the double contour technique. CL, CC, and Real represent the direction of the contour integral. $\Delta$, $\delta$, $R_{cc}$, and $\theta$ represent parameters to determine the shape of the contour. $E_{min}$ is the minimum energy points on the contour and $E_f$ is the Fermi energy.

Finally, we complete an algorithm to calculate self-consistent electron density by adding the effects of the bias voltage as shown in Figure 7.

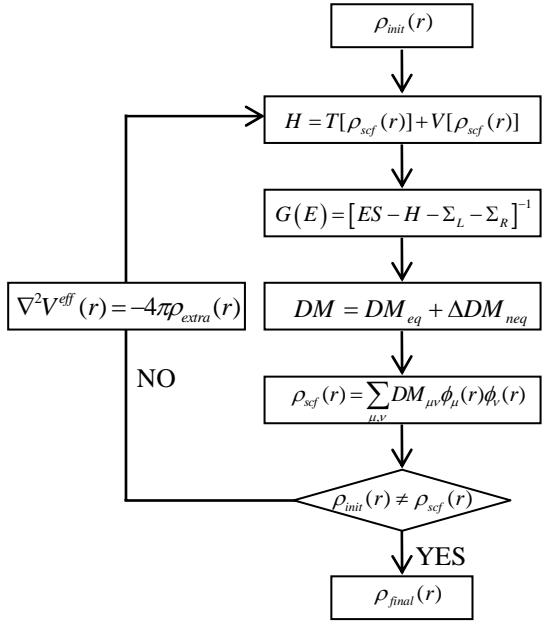

**Figure 7.** Algorithm of the DFT coupled with NEGF to obtain self-consistent electron density under non-equilibrium conditions.

**E. Analysis**

The Green's function with the self-consistent electron density gives information about physical properties of the system as follows. In the first step, the spectral functions are obtained by using Eqs. 13 and 14. Then, the density of states (DOS) for the L-B-D-B-R system can be directly calculated by performing the trace of the spectral functions multiplied by the overlap matrix as

$$DOS(E,V) = \int_{IBZ} w_{k_\parallel} dk_\parallel Tr\left[ A^{k_\parallel}(E,V) S^{k_\parallel} \right]$$

where $w_{k_\parallel}$ is a weighting factor of the lattice vector point $k_\parallel$.

In turn, the transmission function which gives the probability of the electron transport through each energy state can be obtained by using the following equation.

$$T(E,V) = \int_{IBZ} w_{k_\parallel} dk_\parallel T^{k_\parallel}(E,V) \qquad (15)$$

where $T^{k_\parallel}(E,V) = Tr\left[\Gamma_L^{k_\parallel}(E,V) G^{k_\parallel}(E,V) \Gamma_R^{k_\parallel}(E,V) G^{k_\parallel\dagger}(E,V)\right]$.

All variables in Eq. 15 are the function of bias voltage $V$ as well as energy $E$ because we obtain those values from the self-consistent electron density at each contour point under the external bias voltage. The transmission function in Eq. 13 can be related to the transmission matrices in the scattering theory by changing the form slightly. Using the cyclic property of the trace, we obtain the following equation.

$$T^{k_\parallel}(E,V) = Tr\left[t^{k_\parallel}(E,V) t^{k_\parallel\dagger}(E,V)\right]$$

where we define

$$t^{k_\parallel}(E,V) = \Gamma_L^{k_\parallel 1/2}(E,V) G^{k_\parallel}(E,V) \Gamma_R^{k_\parallel 1/2}(E,V).$$

Finally, the corresponding steady current ($I$) can be calculated by using Landauer-Büttiker formula:[55]

$$I(E,V) = \frac{2e}{h} \int_{-\infty}^{\infty} T(E,V) \left[f(E-\mu_L) - f(E-\mu_R)\right] dE$$

where $e$ is the charge of the electron and $h$ is the Plank constant.

## III. Application

### A. Quantum Conductance of a Carbon Nanotube

As the first example of applications, we investigated the quantum conductance phenomena for systems having quantum point contacts. Since CNTs have the ballistic transport phenomena through a long range,[19] they are useful examples for the study of quantum conductance, and can be compared with the same size of graphene sheet which will be discussed in the next section. We choose an armchair CNT with chiral vector (5, 5) [CNT55] with/without nitrogen (N) or boron (B) dopants as shown in Figure 8.

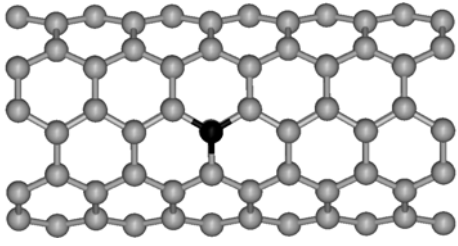

**Figure 8.** CNT55 with N or B dopants. The gray color atoms are carbon and the black one is N or B.

For a clean armchair CNT, its conductance curve shows the clear stepwise structure because each conducting channel corresponding to each band in the electronic structure has the maximum conductance (the unit of the quantum conductance, $G_o=2e^2/h$) due to the quantum point contacts. Figure 9(a) shows the band structure of the CNT55. Delocalized $\pi$ and $\pi^*$ bands cross the Fermi level on the point corresponding to 2/3 between $\Gamma$ (0 k) and X (1 k). These two bands determine the conductance behavior of the CNT55 around the Fermi energy. Each band contributes $1G_o$ in the ballistic transport regime [Figure 9(b)]. Therefore, the conductance value is $2G_o$ around the Fermi level corresponding to the number of the bands on a given energy. In contrast, B/N-doped CNTs do not show such a stepwise conductance structure. They have localized states due to the positive/negative doping effects at lower/higher energy than the Fermi energy as shown in Figure 9(c). These localized states induce scattering potential; hence, the resistance of each conducting channel increases. In particular, the conductance is significantly suppressed at the energy corresponding to the localized states as shown in Figure 9(b), which is consistent with the previous work.[56]

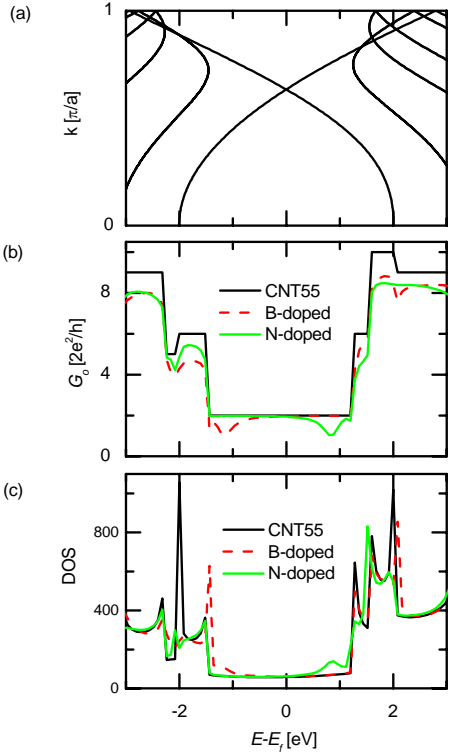

**Figure 9.** Electronic structure of the CNT55. (a) The band structures along the periodic direction. 'a' is the lattice constant of the primitive unit cell. (b) The conductance and (c) DOS for bare and doped CNT55. $G_o$ is the unit of the quantum conductance.

## B. Spin-Polarized Conductance of Graphene Nano-Ribbon

As the second example, we have investigated the transport phenomena of a zig-zag graphene nano-ribbon (ZGNR), as depicted in Figure 10. Four atomic lines including 40 carbon atoms and 4 hydrogen atoms are considered as the unit cell of both electrode parts. The band structures of ZGNRs have been studied in a variety of width recently.[44, 57] They have anti-ferromagnetic ground states in the absence of external electric fields but become ferromagnetic states under the applied fields. In this example, we discuss spin-dependent transport phenomena of the 10-ZGNR (Figure 10) controlled by electric fields.

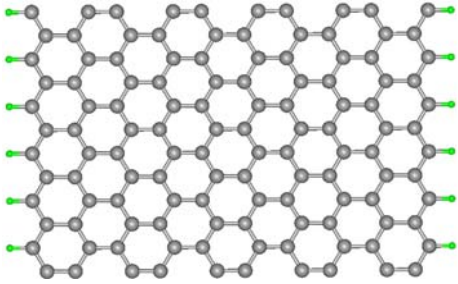

**Figure 10.** A 10-ZGNR structure in a unit cell for the given periodic condition along the periodic direction. All carbon atoms at the left and right edges are bonded with hydrogen atoms. "10" of 10-ZGNR is the usual label related to the width of the nano-ribbon.

If we roll 10-ZGNR along the width, it exactly matches CNT55. However, its band structure is completely different from that of the CNT55 because of the edge effect.[44, 58] In Figure 11(a), which shows the band structure of 10-ZGNR for the spin-unpolarized calculation, the π and π* states make flat band from the 2/3 point to X point instead of crossing the Fermi level, as in the CNT55. The interaction between spin states localized at both edges splits the flat band, and the system becomes anti-ferromagnetic ground state as shown in Figure 11(b). The localized spin states at both edges are differently influenced by external electric fields applied along the direction across the width of ZGNR. Therefore, it becomes a ferromagnetic state, because the spin degeneracy is broken [Figure 11(c)]. We have performed the calculation of conductance corresponding to each band structure in Figure 11(a-c). Integer value of quantum conductance on the given energy point equals to the number of bands on the same energy as shown in Figure 11(d-f), because of the quantum point contacts. Figure 11(e) shows spin dependent conductance for the anti-ferromagnetic state. In fact, both conductance curves of spin-up and spin-down perfectly match. However, as the transverse electric field (0.5 V/Å) is applied, the spin dependent conductance appears around the Fermi level [Figure 11(f)] because of the broken spin degeneracy in the band structure [Figure 11(c)].

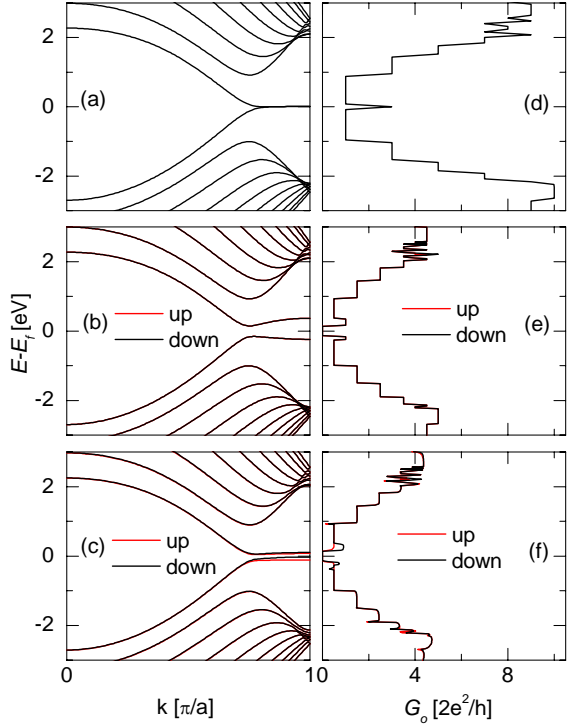

**Figure 11.** Band structures of the 10-ZGNR in the cases of (a) spin-unpolarized, (b) spin-polarized, and (c) spin-polarized with the external electric field calculations and (d-f) the corresponding conductance curves. The applied electric field is applied along the direction to cross the width of the ZGNR as much as $0.5 V/\text{Å}$.

Thus, ZGNRs have unique spin dependent electronic properties. This useful phenomenon coupled with a robust graphene structure can be utilized to develop spintronic devices. Our code would be helpful to understand and predict diverse phenomena arising from such devices.

## C. Conductance of Gold and Nickel nanowires

The simplest 1-dimensional (1D) nanostructure is the perfect linear atomic chain (LC). The third example is quantum conductance for Au and Ni NWs as shown in Figure 12. We use 12 atoms in one unit cell as the L-B-D-B-R system. They also show the stepwise quantum conductance [Figure 12(a,b)]. Au LC has one s-character band and two d-character bands at the Fermi level, though the Au bulk crystal (face centered cubic, FCC) has only one s-character band around the Fermi level.[59] Our calculation accurately describes the quantum conductance phenomena for the corresponding band structures. In Figure 12(a), the conductance values at the Fermi level for Au LC have $3G_o$ exactly. Ni bulk structure (FCC) has value of magnetic moment ($\mu$) of 0.67 $\mu_B$ ($\mu_B$: Bohr magneton) in the experimental measurement and the prediction using the full potential linear augmented plane wave method within DFT shows consistent value (0.61 $\mu_b$).[60] In addition, the theoretical calculation predicts that $\mu$ for Ni increases as the dimensionality decreases, e.g. 1D Ni LC has 1.10 $\mu_b$. Our calculation result has reasonable value, (1.32 $\mu_b$) for Ni LC with the singe-zeta basis size and GGA functional. As a result, it shows spin-polarized conductance curves as shown in Figure 12(b).

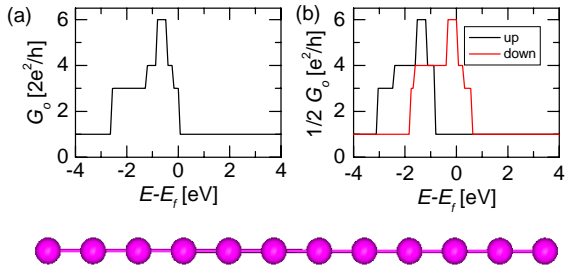

**Figure 12.** Conductance curves of (a) Au and (b) Ni nanowires. The figure below the graphs represents the structure of linear atomic chains which are used in the calculations.

In the active research for NWs of transition metals, identification of their structures is difficult for very thin systems. In experiment, because conductance values depend highly on the delicate structures of the NWs,[61] conductance measurement is used to identify the structures of NWs. Theoretical predictions can give accurate conductance values for given NWs. Comparison between theoretical and experimental results would be very helpful in determining the structures of NWs.[62-65]

**D. Dithiol-Benzene Contacted between Au Bulk Electrodes**

In experiments of molecular electronics, metal-molecule junctions have been formed by self-assembled monolayers.[2, 5, 9, 10] In that case, a molecule is contacted on infinite surface of electrode materials. Therefore, accurate description for Bloch wavefunctions defined in an infinite surface is required in a theoretical study about devices fabricated by self-assembled monolayers. As the last examples, we emphasize how k-point sampling in a calculation of electron transport is important by applying it to the dithiol-benzene molecule which has been most actively investigated experimentally and theoretically.[9, 31, 66, 67] Figure 13 shows geometrical structure of the molecular device which we used in calculations. Three layers at both ends are left and right electrodes, respectively. Periodic condition along a surface-parallel direction is implemented via k-point sampling. Sing-zeta polarization for Au and double-zeta polarization for the others were used as basis orbitals.

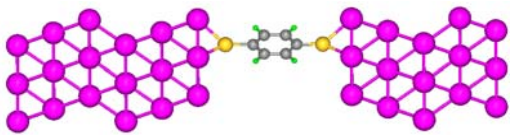

**Figure 13.** Geometry of ditiol-benzene molecule contacted between Au electrodes. The last three layers at both ends are left and right electrodes, respectively.

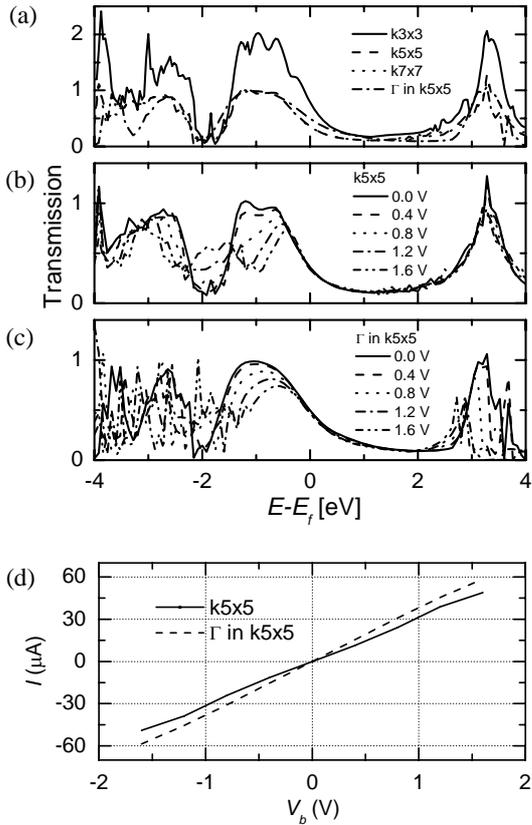

**Figure 14.** Transmission curves of dithiol-benzene molecule: (a) different k-point samplings, (b) different bias voltages with k5x5 and (c) different bias voltages with gamma in k5x5. (d) *I-V* curve corresponding to the transmission curves in (b) and (c).

From Fig. 14(a), note that at least 5x5 k-points in the BZ along the surface-parallel direction is necessary to obtain converged transmission. The transmission curve on the gamma point [$T^{\Gamma}(E)$] in calculation with 5x5 k-points has similar pattern to the total transmission curve [$T(E)$] which is obtained by averaging transmission coefficients in each k-points as described in Eq. 15. This shows that the main origin of such a convergence comes from converging density matrix in accordance of increased number of k-points. Slight discrepancy of them can be understood by considering contribution of incoming Bloch states away from the gamma point in the BZ. In previous works for the same system, all of them did not considered k-point sampling.[31, 66-68] The previous work which was done by the TransSIESTA code shows similar results to our gamma point calculation,[66] but we improved it with the correction due to contribution from the Block states away from the gamma point.

In order to investigate the current-voltage (*I-V*) characteristics, we calculated transmission coefficients at various bias voltages with 5x5 k-points and gamma point as shown in Fig. 14(b) and (c). At the Fermi energy, there is no significant change, but a peak gradually reduces at a lower energy as the voltage increases. Although the changes in $T^{\Gamma}(E)$ and $T(E)$ show similar trends, the peak when $E-E_f$ is around −1 eV is more broadened for $T^{\Gamma}(E)$. The difference in the transmission coefficient is directly reflected in the current-voltage (*I-V*) characteristics. The current increases linearly due to the finite transmission coefficient near the Fermi energy for both cases as shown in Fig. 14(d). In case of $T^{\Gamma}(E, V)$, the *I-V* characteristics is almost similar to that of the previous work.[66] However, in the case of $T(E, V)$, the magnitude of current is smaller because a more significant reduction of height for the peaks around $E - E_f = -1$ eV suppresses the current increase. Thus, the k-point sampling plays an important role in the converged density matrix because of the contribution by Bloch wavefunctions away from the gamma point, which is important for the accurate description of the surface property of electrodes.

**IV. Concluding remarks**

The rapid growth of nanoscience accelerates the study of the nanoelectronics and the efforts to better understand the transport phenomena of nanomaterials have led to the development of various kinds of sophisticated methods. In this regards, we have discussed the first principles method to describe coherent transport phenomena for open systems which are composed of a scattering region and two semi-infinite electrodes. The screening approximation arising from the metallic property of the electrodes allows for an efficient means to handle if they were finite by introducing of the self-energy. The non-equilibrium Green's function method was implemented for use with DFT by using atomic orbital basis sets in the SIESTA code. Our scheme gives self-consistent electron density under the non-equilibrium condition due to external finite bias voltages. We assumed the steady state for the current induced from the bias voltage, so that the exchange-correlation energy of electrons on the scattering region can be described by the static functional (LDA or GGA). As practical applications for this scheme, we have applied it to the investigation of several nano-scaled devices, e.g. nanowires, carbon nantubes, graphene nano-ribbons, and single molecules. It shows clear stepwise conductance curves for systems having quantum point contacts. The spin-polarized conductance for the zig-zag graphene nano-ribbon and ferromagnetic nanowires are accurately described. In this regard, the present theoretical tool could play a potentially useful role in the study of spintronics. In molecular electronics, the metal-molecule junctions, which are mostly formed by self-assembled monolayers in experiments, can be simulated through the realistic modeling of a 2-dimentional infinite surface. In our scheme, we use the k-point sampling technique, which provides improved results compared to the previous calculations of the ditiol-benzene molecule formed by the self-assembled monolayers on the Au(111) surface. We believe that the first principles technique presented in this paper is a useful tool to overcome the limits facing experiments with nano-scaled devices, and it can be easily applicable to other high levels of *ab initio* theory. More importantly, our scheme can readily be extended to describe the time-dependent phenomena and inelastic scatterings.

**Acknowledgment.** This work was supported by GRL(KIKOS) and BK21..